\documentclass[showpacs,preprintnumbers,amsmath,amssymb,aps,twocolumn,superscriptaddress,letterpaper,nofootinbib]{revtex4-1}

\usepackage{amsfonts}

\usepackage{epsfig,amssymb,euscript}
\usepackage{type1cm}
\usepackage{color}

\begin{document}
\title{Strings, Branes, Schwarzian Action and Maximal Chaos}
\author{Avik Banerjee}
\email{avik.banerjee@saha.ac.in}
\author{Arnab Kundu}
\email{arnab.kundu@saha.ac.in}
\affiliation{Theory Division, Saha Institute of Nuclear Physics, HBNI,1/AF Bidhannagar, Kolkata 700064, India.}
\author{Rohan R. Poojary}
\email{ronp@theory.tifr.res.in}
\affiliation{Department of Theoretical Physics, Tata Institute of Fundamental Research, Mumbai 400005, India.}


\begin{abstract} 

In this article, we present explicit evidence that maximal chaos occurs for a generic, probe quark-like defect degrees of freedom, in a strongly coupled large $N_c$ gauge theory. In holography, this corresponds to the dynamics of open string degrees of freedom, in the background of a closed string geometry. In this context, we explicitly show that a Schwarzian effective action for the soft sector emerges and couples with other modes, in the infra-red. This is manifest on an open string worldsheet, as well as a D$1$-brane world-volume, embedded in AdS$_3$. The corresponding maximal chaos is governed by an intrinsic defect D-brane horizon, and an intrinsic non-linear description of the brane or the string. We also present explicit evidence of maximal chaos away from extremality on a D-brane horizon, by computing a four-point out-of-time order correlator of spin-one operators. This further suggests that a similar description of the soft sector physics of open string degrees of freedom may exist in general.

\end{abstract}


\maketitle

\section{Introduction}

Ergodicity and thermalization stand as the cornerstones of understanding dynamical systems, both classically and in the quantum regime. The notion of chaos in such dynamical systems is intimately tied to these ideas. In a quantum mechanical many-body system, the understanding of chaos is currently widely explored, in various systems. Interestingly, by virtue of the gauge-gravity duality\cite{Maldacena:1997re}, for a wide class of such strongly coupled systems these issues have become closely tied to new insights in quantum gravity and conformal field theory, initiated in the early studies in \cite{Shenker:2013pqa, Shenker:2013yza, Shenker:2014cwa}.

In these cases, quantum chaos is defined in terms of out-of-time-ordered correlation (OTOC) functions. Given a pair of Hermitian operators $W(t)$ and $V(0)$, a chaos-diagnostic is defined in terms of the following quantum mechanical observable:
\begin{eqnarray}
C(t) \equiv - \left \langle \left[ W(t), V(0)\right]^2 \right \rangle_{\rm thermal} \ . \label{commutechaos}
\end{eqnarray}
In the classical limit, in which the commutators are replaced by Poisson brackets, the above definition, for position and momentum operators, reduces to the standard notion of classical chaos in terms of exponentially diverging trajectories for infinitesimally close initial conditions, see {\it e.g.}~\cite{Maldacena:2015waa}. This chaotic nature is characterized by the so-called Lyapunov exponent, which can be extracted from a putative exponential growth of the correlator in (\ref{commutechaos}).

The correlator in (\ref{commutechaos}) contains both time-ordered and out-of-time-ordered correlation functions. The former is not sensitive to chaotic behaviour and, effectively, it is sufficient to consider the following correlation instead, to read of the corresponding Lyapunov exponent:
\begin{eqnarray}
{\cal O} \left(t_1, t_2, t_3, t_4 \right) = \frac{\left \langle V(t_1) W(t_2) V(t_3) W(t_4) \right \rangle}{\left \langle V(t_1) V(t_3) \right \rangle \left \langle W(t_2) W(t_4) \right \rangle} \ , \label{chaosdiag}
\end{eqnarray}
where the denominator is chosen for normalization.

Based on analytic arguments, an upper bound of the thus-extracted Lyapunov exponent was derived in \cite{Maldacena:2015waa}: $\lambda_{\rm L} \le \left(2\pi k_{\rm B} T \right)/\hbar$, where $k_{\rm B}$ and $\hbar$ are the Boltzmann and Planck constant, and $T$ is the temperature of the corresponding thermal state. We will work with $\hbar=1=k_{\rm B}$.

The saturation of this bound is certainly a physically interesting corner. So far, several examples of systems are known to saturate this bound: $(i)$ Black holes in Einstein-Hilbert gravity\cite{Shenker:2013pqa, Shenker:2013yza, Shenker:2014cwa}, $(ii)$ conformal field theories with a large central charge and a sparse spectrum\cite{Roberts:2014ifa, Fitzpatrick:2016thx}, $(iii)$ SYK-type models\cite{Maldacena:2016hyu, footnote1} and also $(iv)$ on the induced string worldsheet horizon\cite{Murata:2017rbp, deBoer:2017xdk}. The chaos bound saturation serves as an indication when a system with large degrees of freedom may have a gravitational dual description. In this article, we will demonstrate a simple connection, in terms of a Schwarzian effective action, between the SYK-type models and the open string worldsheet description. Furthermore, we will offer additional evidence of the chaos-bound saturation on D-brane horizons, which is completely intrinsic to the D-brane physics, and therefore corresponds to the dynamics of open strings.

To proceed further, recall that the strongly coupled infra-red(IR) limit of the SYK-type models are described by an effective Schwarzian derivative action, which describes the Goldstone modes due to a spontaneous breaking of the IR conformal symmetry. A similar Schwarzian effective action can be obtained from a two-dimensional Jackiw-Teitelboim (JT) gravity theory\cite{Maldacena:2016upp, Nayak:2018qej}. A subsequent linearised analysis of the effective Schwarzian action leads to the chaos bound saturation for a free probe field coupled to this Schwarzian derivative theory\cite{Maldacena:2016upp}. Sparked by these similarities, very interesting attempts have been made to explicitly construct the string-dual of the SYK-theory, see {\it e.g.}~\cite{Polchinski:2016xgd, Mandal:2017thl, Gross:2017hcz, Das:2017pif, Gaikwad:2018dfc}. Indeed, explicit connection has been established with the two-dimensional quantum gravity model of Polyakov. Also, in \cite{Cai:2017nwk} a similar statement to ours was conjectured. In this article we provide a more direct and explicit evidence of the Schwarzian, including its coupling with other heavy modes in the spectrum.

Since, however, the anomalous dimensions in the SYK-type models are order one numbers, an explicit string dual of the SYK-model is perhaps rather complicated. In this article, motivated by a {\it semi-holographic} philosophy, we make explicit connections between the IR-physics of the SYK-type models, with that of an open string. This is achieved by demonstrating how Nambu-Goto action yields an effective Schwarzian action. By {\it semi-holography}, we simply mean a universal and simple IR physics that may have rather varied and complicated UV-completion. We will not attempt to comment on the UV-physics here. The details of our results will be provided in a companion work\cite{tbp}.

\section{Schwarzian Action from Strings}

Let us begin with the description of a fundamental string in a target AdS$_3$-space, in Fefferman-Graham gauge:
\begin{eqnarray}
ds^2 = - \frac{1}{4} \left( r - \frac{r_{\rm H}}{r}\right)^2  dt^2 + \frac{1}{4} \left( r + \frac{r_{\rm H}}{r}\right)^2  dX^2 + \frac{dr^2}{r^2}  . \label{btzfg}
\end{eqnarray}
The corresponding temperature is: $T = r_{\rm H}/(2\pi)$. The dynamics of the string is described by the Nambu-Goto action:
\begin{eqnarray}
S_{\rm NG} = - \frac{1}{2\pi\alpha'} \int d\tau d\sigma \sqrt{- {\rm det} \gamma}  , \quad \gamma_{ab} = P[g]_{ab}  , \label{nambugoto}
\end{eqnarray}
where $P$ denotes the pull-back operation, $\alpha'$ is the inverse string tension. It is instructive to work in the static gauge: $\tau= t  , \sigma = r $, with the embedding function: $X(t, r)$. As in \cite{deBoer:2017xdk}, a simple embedding is described by $X = 0$, which induces an AdS$_2$(black hole) geometry on the worldsheet, with a Ricci scalar: ${\cal R}_{(2)} = - 2 $. This is precisely the AdS$_2 \subset$AdS$_3$ bisection of the geometry.

Consider any arbitrary solution to the Nambu-Goto(NG) $eom$. Around this classical saddle, denoted by $\bar{X}$, there are small fluctuations. Denoting these fluctuations by $X(\sigma,\tau)$, the linearized equations can be solved, in terms of two undetermined functions:  
\begin{eqnarray}
 X(\sigma,\tau) = \sigma^\Delta \sum_{n=0}^{\infty} \frac{ X^{(n)} (\tau)}{\sigma^{n}} \ . \label{anfg}
\end{eqnarray}
A direct calculation yields: $\Delta = 0, \, -3$, which correspond to the two independent modes: $ X^{(0)}(\tau),  X^{(3)}(\tau)$. Regularity (depending on the physics in question) relates these two modes and therefore the full fluctuations depend on an arbitrary boundary motion of the string $ X^{(0)}(\tau)$.

The Nambu-Goto action has a gauge symmtery pertaining to the worldsheet diffeomorphisms. Since this is a theory defined on a manifold with a boundary, there are certain large gauge transformations which are genuine symmetries of the theory. We claim that the ``soft" modes generated by these symmetries $i.e.$ large diffeomorphisms; are responsible for chaos as observed in \cite{deBoer:2017xdk}. Further we hint at a possible existence of a Schwarzian action associated with these modes.

We do not present a complete worldsheet analysis determining all relevant large diffeomorphisms. In general we expect the large diffemorphisms to depend on the fluctuations (\ref{anfg}). However if one assumes otherwise then the $2d$ large diffeomorphisms inherent to AdS$_2$ structure would be enough. We here try to work with certain large diffeomorphisms which depend on (\ref{anfg}). We use the AdS$_3$ background to our advantage and find a suitable map of the large Brown-Henneaux diffeomorphims in AdS$_3$, projected onto the worldsheet. An independent analysis done completely from the 2d worldsheet perspective should, nonetheless, be possible. 

To see the full effect of the worldsheet diffeomorphisms one would have to relax the static gauge condition. Only then {\it any} worldsheet diffeomorphism can be rewritten as fluctuations of the Nambu-Goto fields $i.e.$ the AdS$_3$ co-ordinates. For convenience, however, we will work in the static gauge. In this gauge, the worldsheet metric for any arbitrary fluctuation of the string is:
\begin{equation}
ds^2 = \frac{dr^2}{r^2} - \frac{1}{4}\left(r-\frac{r_{\rm H}}{r}\right)^2dt ^2+\frac{1}{4}\left(r+\frac{r_{\rm H}}{r}\right)^2dX_{(r,t)}^2,
\end{equation}  
where $r=\sigma\,\&\,t=\tau$. Any worldsheet diffeomorphism will cast this metric into:
\begin{eqnarray}
ds^2 &=&\frac{dr_{(\sigma,\tau)}^2}{r_{(\sigma,\tau)}^2}-\frac{1}{4}\left(r_{(\sigma,\tau)}-\frac{r_{\rm H}}{r_{(\sigma,\tau)}}\right)^2dt_{(\sigma,\tau)}^2\cr
&&+\frac{1}{4}\left(r_{(\sigma,\tau)}+\frac{r_{\rm H}}{r_{(\sigma,\tau)}}\right)^2dX_{(\sigma,\tau)}^2,
\end{eqnarray}
Therefore, for any $X_{(r,t)}$, there exists a soft mode sector associated with it (corresponding to the large gauge transformations). These large diffeomorphisms can be found by projecting the Brown-Henneaux diffeomorphism onto the fluctuating string worldsheet.

The Schwarzian action is most apparent on the soft modes associated with $X_{(r,t)}=0$ solution. In this case the the worldsheet metric is that of AdS$_2$. It can be shown  \cite{Roberts:2012aq} that the most general AdS$_2$ metric:
\begin{equation}
ds^2=\frac{d\sigma^2}{\sigma^2}-\frac{1}{4}\left(\sigma-\frac{L}{\sigma}\right)^2d\tau^2 \label{ads21}
\end{equation}
can be obtained by a finite diffeomorphism from:
\begin{equation}
ds^2 = \frac{dr^2}{r^2}-\frac{1}{4}\left(r-\frac{r_{\rm H}}{r}\right)^2dt^2 \ , \label{ads22}
\end{equation}
using:
\begin{eqnarray}
r=\frac{r_{\rm H}}{R}\left(\sqrt{2}Y+\sqrt{2Y^2-R^2}\right),\,&&e^{2r_{\rm H} t}=\left({Y^2-\frac{R^2}{2}}\right)\cr
Y=f+\frac{2f'^2f''}{r^2f'^2-f''^2},\,&&R=\frac{2\sqrt{2}rf'^3}{r^2f'^2-f''^2} \ , \label{ldiff2}
\end{eqnarray}
where
\begin{equation}
L=\lbrace f,\tau \rbrace=\frac{3f''^2-2f'f'''}{2f'^2}\rightarrow\frac{r_{\rm H}f'^4+3f''^2-2f'f'''}{2f'^2} \ . \label{ldiff2L}
\end{equation}
for $f$ $\rightarrow$ $e^{f\sqrt{r_{\rm H}}}$ which is required to map re-parametrizations of $t$ in terms of boundary time $\tau$ $i.e.$ $t\rightarrow f(\tau)$.  The data in (\ref{ads21})-(\ref{ldiff2L}) define the projection of large diffeomorphisms of AdS$_3$ on the worldsheet. This makes the appearance of the Schwarzian manifest.

Now, we determine how these soft modes couple to arbitrary on-shell fluctuations. As in \cite{deBoer:2017xdk}, our fluctuations and likewise diffeomorphisms are of the order of $\delta=\sqrt{\alpha'}/r_{\rm H}$. This interaction essentially determines the corresponding $4$-point (in general, an $n$-point) correlator in the boundary theory. Generally the large diffeomorphisms may depend on the fluctuations themselves\footnote{The conclusions that we draw, however, do not require this to hold in particular.}. The coupling is determined as follows\footnote{One needn't use the 3d diffeos to construct the 2d diffeos if dependece of diffeos on fluctuations isn't required.}: Consider the $3$d Brown-Henneaux diffeos
\begin{equation}
r\rightarrow \sigma+\delta\xi^\sigma(\sigma,\tau,x),\,\,t\rightarrow \tau+\delta\xi^\tau(\sigma,\tau,x),\,\,X\rightarrow x+ \delta\xi^x(\sigma,\tau,x)
\end{equation} 
The 2d diffeos can be constructed out of these by substituting for $x=x(\sigma,\tau)$ which is a solution in (\ref{anfg}):
\begin{equation}
r\rightarrow \sigma+\delta\xi^\sigma(\sigma,\tau,x_{(\sigma,\tau)}),\,\,t\rightarrow \tau+\delta\xi^\tau(\sigma,\tau,x_{(\sigma,\tau)}) \ .
\end{equation}
The change in the induced metric is then given by the action of a Lie derivative on the metric, along the projected $\xi$-vector. Now, the NG action can be evaluated on-shell. The divergence of the on-shell action can be regulated by adding a boundary counter-term: $\int_\partial d\tau \sqrt{-h}$. Thus, any finite contribution comes solely from the event horizon. Since we plug in fluctuations as a power series expansion in $1/r$ we therefore find the on-shell action as a power series in $\delta$ and $1/r_{\rm H}$.

The Lie derivative is a linear action, the purely Schwarzian term appears in the linearized form, with a $ \epsilon'''(t)$ term, similar to \cite{Nayak:2018qej}. This $\epsilon'''(t)$ function is non-linearly completed to (\ref{ldiff2L}), and occurs at the  order $\delta/r_{\rm H}$. The lowest order coupling of the source $ X^{(0)}(t)$ with the $\epsilon(t)$ modes occurs at the order $r_{\rm H}\delta^3$ and involves a cubic coupling of the form: $\left(\partial_t X^{(0)} \right) \epsilon(t)  \left(\partial_t^2  X^{(0)} \right)$. The dependence on temperature $\beta=2\pi/r_{\rm H}$ of this coupling can be shown to be consistent with the scrambling time, as obtained in \cite{deBoer:2017xdk}. Note that, the non-linearity of the underlying theory, be it Einstein-gravity in general, and a Nambu-Goto action in our case, lies at the core of such a coupling. Thus, while a linearized analysis can already identify the effective Schwarzian sector (in general, the soft sector), the coupling of this soft sector with generic fluctuations are determined at a non-linear order. 
\begin{figure}[h!]
\centering
\includegraphics[scale=0.25]{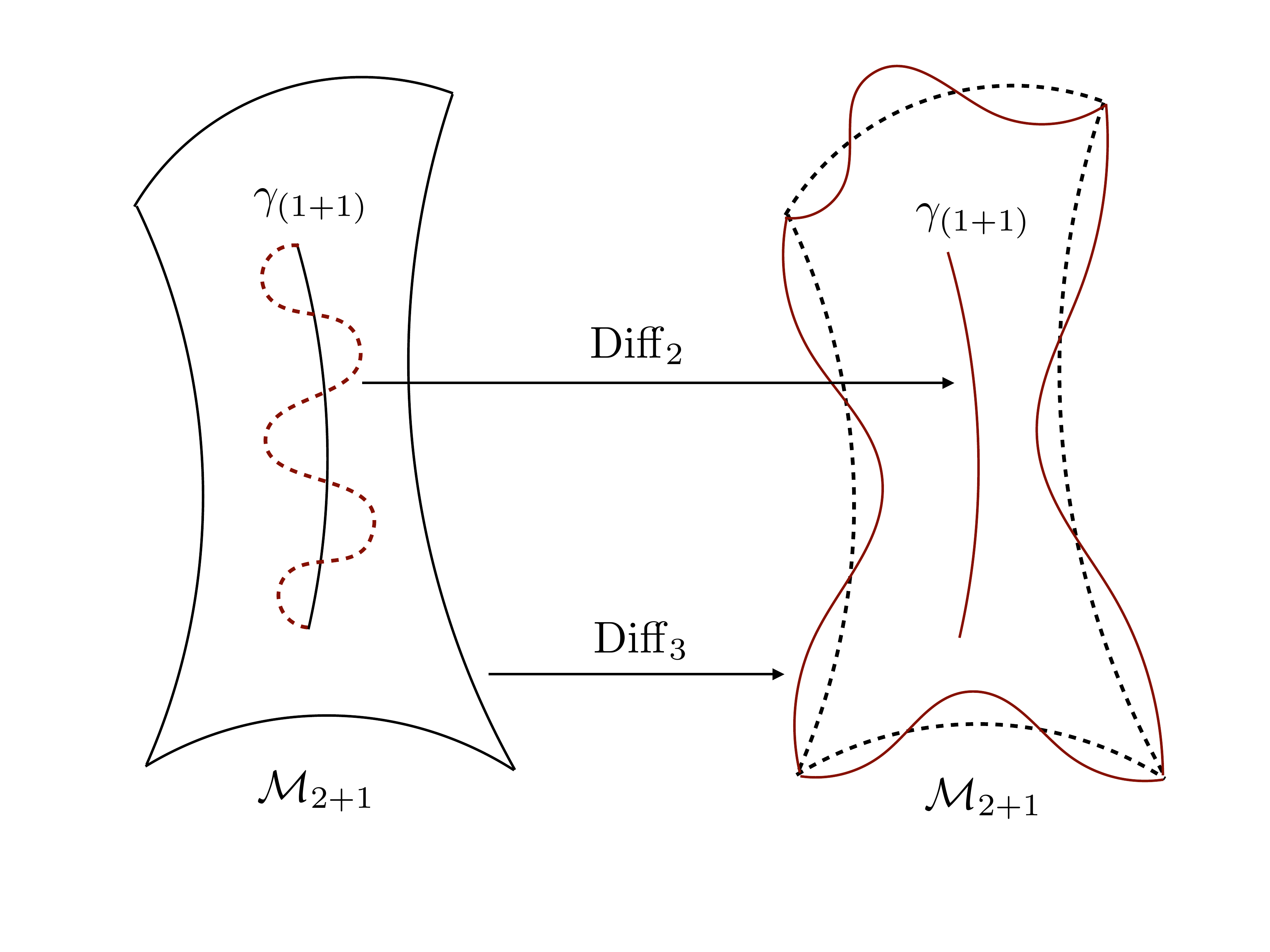}
\caption{\small The $(1+1)$-dimensional geometry, embedded in a $(2+1)$-dimensional manifold. The dashed curve on the left represents fluctuations around the classical embedding, which, on-shell, is mapped to the classical embedding with a ${\rm Diff}_2$ transformation. This ${\rm Diff}_2$ is obtained from the ${\rm Diff}_3$ transformation of the embedding space ${\cal M}_{2+1}$. Here, ${\cal M}_{2+1} \equiv {\rm AdS}_3$.}
\label{stringdiffeo}
\end{figure}

Consider now a simple example, in which we take $r_{\rm H} =0$ in (\ref{btzfg}). Now we impose a non-trivial boundary condition on the string end point: it undergoes an uniform acceleration. The corresponding classical embedding function is non-trivial and is given by\cite{Xiao:2008nr, Jensen:2013ora}
\begin{eqnarray}
X^{(0)} \left(t, z \right) = \pm \sqrt {a^{-2} + t^2 - z^2 } \ , \label{classacc}
\end{eqnarray}
where $a$ is the magnitude of the constant acceleration. The $\pm$ sign can be physically attributed to the identification of a quark or an anti-quark as the end point of the string. The induced metric is an AdS$_2$-black hole. Now, one can analyse transverse fluctuations, and viewing them as a change in the worldsheet metric, one obtains: $ \left. \delta {\cal R}_{(2)} \right|_{\rm on-shell}=  0 $. Thus, evidently, these small fluctuations can be viewed as worldsheet diffeomorphism. Proceeding as before, we can therefore readily identify a large diffeomorphism in (\ref{ldiff2})-(\ref{ldiff2L}). Also note that, even though we have set $r_{\rm H} =0$, due to the uniform acceleration, the worldsheet induced AdS$_2$ metric inherits an event horizon.

Finally, note that the most general solution of the string embedding function, subject to the most general boundary condition is obtained by Mikhailov\cite{Mikhailov:2003er}. This is described by two arbitrary smooth functions, constrained by one relation. Physically, this embedding represents a string whose end-points undergo an arbitrary time-like trajectory at the conformal boundary. It is easy to check that the induced worldsheet metric yields ${\cal R}_{(2)} = - 2 $. Uniformization theorem ensures that any string profile with a time-like boundary condition at the end point is diffeomorphic to the Mikhailov embedding. Thus, for any physical embedding, our previous analyses of the large diffeomorphisms and their corresponding coupling with other fluctuation modes on the worldsheet holds true.

Therefore, the effective action for the soft modes, in its non-linear form is simply obtained to be:
\begin{eqnarray}
S_{\rm NG}^{({\rm ren})} = \frac{\epsilon_{\rm IR}}{4 \pi \alpha'} \int dt \left\{f(t) , t \right\} \ , \label{effSchwarzian}
\end{eqnarray}
where $\epsilon_{\rm IR}$ is the infrared cut-off. A physically relevant identification of $\epsilon_{\rm IR}$ is the event horizon itself\footnote{In our case $\epsilon_{\rm IR}\equiv \sqrt{\alpha'} \beta^2$.}. This is precisely the Schwarzian action that governs the IR-dynamics. Keeping the explicit cut-off surface breaks conformal invariance spontaneously, which allows us to extract the Schwarzian effective action. The maximal chaos saturation, showed in \cite{Murata:2017rbp, deBoer:2017xdk}, is therefore observed in the $4$-point OTOC of modes that couple with the Schwarzian soft sector. This applies generally and the corresponding chaos-bound saturation should hold beyond primary operators.

\section{Schwarzian Action from Brane}

It is rather straightforward to generalize the above analyses for a D$1$-brane, embedded in an AdS$_3$ in a similar way. This background AdS$_3$ may simply originate from a standard D$1$-D$5$ brane system. The only difference will be in the pre-factor in (\ref{effSchwarzian}): instead of the string tension, D$1$-brane tension will appear. This pre-factor determines, {\it e.g.}~the scrambling time for the degrees of freedom living on the brane, similar to \cite{deBoer:2017xdk}. The $4$-point function ${\cal O}(t)$, defined in (\ref{chaosdiag}), behaves as: ${\cal O}(t) = 1 - \alpha T_{{\rm D}1}^{-1} \left( e^{\lambda_{\rm L} t} \right)$, where $\alpha$ is a purely numerical quantity and $T_{{\rm D}1}$ is the tension of the D$1$-brane.  The scrambling time, denoted by $t_{\rm sc}$, is defined as: $\alpha T_{{\rm D}1}^{-1} \left( e^{\lambda_{\rm L} t_{\rm sc}} \right) \sim {\cal O}(1)$. D$1$-brane tension is given by $T_{{\rm D}1} \sim 1/(g_{\rm s} \alpha' )$, where $g_{\rm s}$ is the string coupling. Gauge-gravity dictionary contains: $\alpha' \sim 1/(\sqrt{\lambda})$ and $g_{\rm s} \sim \lambda/N_c$, where $N_c$ is the rank of the gauge group and $\lambda$ is the corresponding 't Hooft coupling. Thus, the corresponding scrambling time is given by $t_{\rm sc} \sim \beta \log \left( \frac{\sqrt{\lambda}}{N_c}\right)$, which is different from the scrambling time of a Einstein-gravity black hole, given by $t_{\rm sc} \sim \beta \log N_c^2$\cite{Sekino:2008he, Lashkari:2011yi}, and the scrambling time on a string worldsheet, given by $t_{\rm sc} \sim \beta \log \sqrt{\lambda}$\cite{Murata:2017rbp, deBoer:2017xdk}. In fact, the D$1$-brane scrambling time is further parametrically suppressed compared to the string worldsheet result.

Moreover, one can consider a broader class of examples with a D$1$-brane, following {\it e.g.}~\cite{Das:2010yw}: a rotating D$1$-brane in an AdS$_5 \times S^5$ background. The rotation induces a worldvolume horizon for the D$1$-probe sector and, correspondingly, the zero dimensional defect hypermultiplets couple to the ${\cal N}=4$ super Yang-Mills with a time-dependent coupling. The fluctuation sector, once again, acts as the diffeomorphism on the worldvolume AdS$_2$, on-shell. Thus, one obtains an effective Schwarzian action from the D$1$-brane description. This is perhaps unsurprising since S-duality maps fundamental strings to D$1$-branes. Note, however, not all physics are invariant under this map, {\it e.g.}~the scrambling time is a sensitive function of this.

\section{Maximal Chaos on a Brane Horizon}

In this section, we will consider the non-linear description on a D-brane and the implicit presence of open string degrees of freedom which also results in maximal chaos, away from the extremal limit. Let us consider the brane configuration corresponding to a defect CFT which is localized on an ${\mathbb R}^{2,1} \subset {\mathbb R}^{3,1}$, following \cite{DeWolfe:2001pq, Sonner:2012if}. The defect CFT is coupled to the ${\cal N}=4$ SYM, which is defined on the ${\mathbb R}^{3,1}$. The geometric representation of this construction is given by an $N_f$ number of probe D$5$-branes placed in an AdS$_5\times S^5$ geometry. In the probe limit, we have $N_c \gg N_f$, where $N_c$ is the number of D$3$-branes.  

Explicitly, in a co-ordinate patch, the configuration is summarized in table \ref{d3d5}.
\begin{table}[h!]
	\begin{center}
  			    \begin{tabular}{l|c|c|c|c|c|c|c|c|c|c|r}
			\textbf{} & $r$ & $t$ & $x^1$  & $ x^2 $ & $ x^3 $ & $ \psi $ & $ \theta $ & $ \phi $ & $ \chi $ & $ \zeta $\\ 
		
			\hline
			D3 & $\times$ & $\checkmark$ & $\checkmark$ & $\checkmark$ & $\checkmark$ & $\times$ & $\times$ & $\times$ & $\times$ & $\times$\\ 
			D5 & $\checkmark$ & $\checkmark$ & $\checkmark$ & $\checkmark$ & $\times$ & $\times$ & $\checkmark$ & $\checkmark$ & $\times$ & $\times$\\ 
			
		\end{tabular} 
		\caption{\small The AdS$_5$ is covered by the co-ordinates $\{t, x^i, r\}$ and the $S^5$ is covered by $\{\psi, \theta, \phi, \chi, \zeta\}$. The D$5$-branes wraps an $S^2\subset S^5$, which is covered by $\{\theta, \phi \}$.}
		\label{d3d5}
	\end{center}
\end{table}
We work with a standard Poincar\'{e} metric in AdS$_5$, which has no event-horizon and therefore no thermal physics. The dynamics of the probe D$5$-sector is governed by a Dirac-Born-Infeld (DBI) action:
\begin{eqnarray}
S_{\rm DBI} = - T_{{\rm D}5} \int d^6\xi \sqrt{- {\rm det} \left( P[g] + \left( 2 \pi \alpha' \right) F \right)} \ ,
\end{eqnarray}
where $F$ is the U$(1)$ gauge field on the brane. 

We pick a particular brane embedding, characterized by $\psi= \psi(r)$ and the gauge field $A = \left( - E t - a_x(r) \right) dx^1$. Physically, this corresponds to coupling the defect fundamental degrees of freedom to a constant electric field. This field pair creates and drives a matter current, see {\it e.g.}~\cite{Karch:2007pd, Albash:2007bq, Erdmenger:2007bn, Bergman:2008sg, Johnson:2008vna, Alam:2012fw}. All subsequent fluctuations of the D$5$-brane couple to the corresponding open string metric (osm), given by\cite{Seiberg:1999vs}: ${\cal S}_{ab} = P[g]_{ab} - \left( F \cdot P[g]^{-1} \cdot F \right)_{ab} \ .$ Because of the non-vanishing electric field, the osm has an event-horizon, and due to a corresponding open string equivalence principle\cite{Kundu:2013eba, Kundu:2015qda, Banerjee:2015cvy, Banerjee:2016qeu}, all fluctuations have an effective thermal description, at an effective temperature $T_{\rm eff} \sim \sqrt{E}$.

On this classical profile, consider gauge field fluctuations, similar to the ones considered in \cite{Sonner:2012if}:
\begin{equation}
\delta A =  \delta A_{\perp}(t,r) ~dx^2 \ , \label{gaugefluc}
\end{equation}
Our goal is to compute a current-current $4$-point OTOC, comprising of the fluctuations $\delta A_\perp$. This correlation function can be recast as a $2-2$ scattering amplitude calculation in the corresponding Penrose diagram\cite{Shenker:2014cwa, deBoer:2017xdk}. The result of this calculation is\cite{tbp}:
\begin{eqnarray}
{\cal O} \left( t+ i \frac{\beta}{4}, i \frac{\beta}{2} , t - i \frac{\beta}{4}, 0 \right) = 1 -  \frac{ C e^{\frac{3 \pi}{2}}}{8 r_{\rm E}^2  ~ T_{{\rm D}5} } e^{\frac{2 \pi}{\beta} t} \ ,
\end{eqnarray}
where $C$ is a numerical constant, $r_{\rm E}^4 = E^2$. The corresponding Lyapunov exponent therefore saturates the bound: $\lambda_{\rm L} = \frac{2\pi }{\beta} = 2\pi T_{\rm eff} $.

Note that, the D$3$-brane modes, which give rise to the gravity in AdS$_5\times S^5$ do not have any exponentially growing mode, for those are kept at a vanishing temperature. Thus, there are two decoupled sectors, one exhibiting maximal chaos due to the probe D-brane horizon and the other having a vanishing Lyapunov exponent. In fact, if we consider an AdS$_5$-Schwarzschild background, corresponding to a temperature $T$, the gravity modes will saturate the bound $\lambda_{\rm L}^{\rm gravity} = 2 \pi T$ while the brane modes will saturate $\lambda_{\rm L}^{\rm brane} = 2 \pi \left( T^4 + E^2  \right)^{1/4} > \lambda_{\rm L}^{\rm gravity}$. This is true only at the probe limit. 

Finally, the scrambling time on the D$5$-brane fluctuations are given by $t_{\rm sc} \sim \beta \log \left( N_c N_f\sqrt{\lambda}\right) $. This is clearly different from the result on the string worldsheet, as well as the D$1$-brane. Motivated by these observations, we conjecture a simple generalization of the scrambling time on a D$p$-brane:
\begin{eqnarray}
t_{\rm sc} \sim \beta \log \left| \frac{\lambda^{(3-p)/4}}{N_c N_f}\right| \ , \label{scramblegen}
\end{eqnarray}
which is obtained by converting the tension of a D$p$-brane in to the dual gauge theory parameters. Thus, the brane horizon scrambling can be even further suppressed, provided $p<3$. Thus, in the probe limit, the brane horizon can be a faster scrambler and yield a larger Lyapunov exponent, compared to the background gravity geometry. For $p=3$, we recover the gravity result, with an identification $N_f \sim N_c$.

\section{Generalized Soft Sector \& Chaos Bound}

Given the ubiquitous appearance of the Schwarzian derivative function, it is natural to think about an infra-red sector which is described by a general functional of the Schwarzian derivative, which still preserves the SL$(2)$ global symmetry. A simple class of examples is given by the Schwarzian derivative, raised to a power $n$. For a given $n$, we can derive the classical equation of motion and subsequently determine the propagator. In the frequency space, this propagator has multiple zero modes, located at $\omega = 0 , \pm i, \pm i \frac{\sqrt{2n-1}}{\sqrt{2n-2}}$. For $n=1$ (which needs to be treated separately), {\it i.e.}~the standard Schwarzian action, these zero modes are located at $\omega = 0 , \pm i$.  

One can now calculate the propagator in time-space, and using that, calculate ${\cal O} \left( t_1, t_2, t_3, t_4 \right)$, with $t_1>t_3>t_2>t_4$, following \cite{Maldacena:2016upp}. The corresponding OTOC, when analytically continued to real time, subsequently yields an exponential growing mode in which the Lyapunov exponent saturates the maximal bound\cite{tbp}: $\lambda_{\rm L} = 2\pi T \kappa_n$, where $\kappa_n = {\rm max}(1, \frac{\sqrt{2n-1}}{\sqrt{2n-2}})$. This exponential growth is dominated by the largest zero mode, and can therefore violate the chaos-bound. The result above mimics a similar bound in theories with higher spin degrees of freedom discussed in \cite{Perlmutter:2016pkf}. Based on general grounds, and imposing the chaos-constraint, the allowed range is $n \in (-\infty, 0) \cup (0, 1]$. Incidentally all negative values of $n$ are also allowed. For all of these cases, one obtains a maximal chaos. 

Furthermore, the information of this exponent $n$ enters in determining the corresponding scrambling time, which is given by $t_{\rm sc} \sim \beta \log {\cal K}(n) $, where ${\cal K}(n)$ is a function of $n$. The allowed range of $n$ yields a standard scrambling time, while the chaos-bound violating regime yields a {\it too-fast} scrambling. This is also similar to the observation in \cite{Perlmutter:2016pkf}.

\section{Conclusions}

In this article, we have made a simple explicit demonstration of how an effective Schwarzian action can emerge from a Nambu-Goto string worldsheet, and how the Schwarzian sector couples with other fluctuation modes on the string. This argument holds true for D$1$-branes as well. Furthermore, we have presented an explicit example of a defect CFT, in terms of a D$3$-D$5$ brane system and computed explicitly a $4$-point OTOC comprising of current-current correlation function. This correlator also exhibits a saturation of maximal chaos, in which the temperature is determined by the brane horizon. This brane horizon is completely independent of the gravity horizon and thus corresponds to an intrinsic, non-linear physics on the brane. Furthermore, we have presented a conjectured formula for scrambling time on such defect CFT systems, which can exhibit a faster scrambling compared to the usual large central charge CFTs with an Einstein-gravity dual.

To the best of our knowledge, this is the first calculation of a $4$-point OTOC involving spin-$1$ operators, which is subsequently saturating the chaos-bound. This indicates a stronger universality of the IR-physics. It also hints that an universal Schwarzian soft sector may be present in a wide class of D-brane systems, provided an AdS$_2$ submanifold is realized. We have obtained this Schwarzian on the string worldsheet and the D$1$-brane worldvolume, by considering those as embedded hypersurfaces in an asymptotically AdS$_3$-background. The Schwarzian sector originated from this background AdS$_3$. However, we have observed the ubiquity of the chaos-bound saturation on string and brane fluctuations in a general AdS$_{d+1}$-background. Also, it is reasonable to expect that the string worldsheet has an intrinsic description, without making any reference to an embedding, it is natural to assume that an effective Schwarzian action will indeed emerge from an open string degree of freedom, in a background-independent manner. The feature of maximal chaos holds true for near-extremal and non-extremal horizons. In view of this, it is reasonable to expect an effective Rindler description of the corresponding soft modes, for a general horizon, which is a very intriguing possibility to explore.

We find various other interesting questions to explore further. The precise mechanism behind the emergence of the light modes from an open string, a detailed analysis of how it couples to the massive modes in the spectrum, the precise relation between the worldsheet CFT and the spacetime CFT, and an understanding of the chaos-bound saturation from the perspective of a defect CFT would be some of the very interesting questions to further address. Finally, a more explicit connection with the SYK-type models, from a {\it semi-holographic} point of view or otherwise, would be a very interesting direction to explore.

\section{Acknowledgements}

We thank Jan de Boer, Gautam Mandal, Juan F.~Pedraza, Shibaji Roy and Harvendra Singh for many illuminating conversations and discussions. We thank Augniva Ray for collaboration on related matters. We specially thank Gautam Mandal and Juan F.~Pedraza for patiently explaining many issues, related to the chaos-bound computation, to us. AK and RP acknowledge the support by the International Centre for Theoretical Sciences (ICTS) for local hospitalities and providing a stimulating environment during the program ``AdS/CFT at $20$ and Beyond," where a part of this work was done. We also thank DAE, Government of India for support.

\bibliographystyle{apsrev}


\end{document}